\documentclass[aps,prb,twocolumn,groupedaddress]{revtex4-1}

\usepackage{graphicx,color,graphics}
\usepackage{dcolumn} 
\usepackage{amsmath}
\usepackage{amsfonts}
\usepackage{amssymb}
\usepackage{hyperref}
\usepackage{url}
\usepackage{bm}
\usepackage{xcolor}

\newcommand*{\citen}[1]{%
  \begingroup
    \romannumeral-`\x 
    \setcitestyle{numbers}%
    \cite{#1}%
  \endgroup   
}

\begin{document}

\title{Multi-Element Logic Gates for Trapped-Ion Qubits}

\author{T. R. Tan}
\author{J. P. Gaebler}
\author{Y. Lin}
\altaffiliation{Current address: JILA, University of Colorado and National Institute of Standards and Technology/Department of Physics, University of Colorado, Boulder, Colorado 80309, USA.}
\author{Y. Wan}
\author{R. Bowler}
\altaffiliation{Current address: University of Washington, Department of Physics, Box 351560, Seattle, Washington 98195, USA.}
\author{D. Leibfried}
\author{D. J. Wineland}
\affiliation{National Institute of Standards and Technology, 325 Broadway, Boulder, Colorado 80305, USA.}

\unitlength 1in

\date{\today}

\maketitle

{\bf Precision control over hybrid physical systems at the quantum level is important for the realization of many quantum-based technologies. In the field of quantum information processing (QIP) and quantum networking, various proposals discuss the possibility of hybrid architectures \cite{Wallquist2009} where specific tasks are delegated to the most suitable subsystem. For example, in quantum networks, it may be advantageous to transfer information from a subsystem that has good memory properties to another subsystem that is more efficient at transporting information between nodes in the network. For trapped-ions, a hybrid system formed of different species introduces extra degrees of freedom that can be exploited to expand and refine the control of the system. Ions of different elements have previously been used in QIP experiments for sympathetic cooling \cite{Barrett2003}, creation of entanglement through dissipation \cite{Lin2013}, and quantum non-demolition (QND) measurement of one species with another \cite{Hume2007}. Here, we demonstrate an entangling quantum gate between ions of different elements which can serve as an important building block of QIP, quantum networking, precision spectroscopy, metrology, and quantum simulation. A geometric phase gate between a $^9$Be$^+$ ion and a $^{25}$Mg$^+$ ion is realized through an effective spin-spin interaction generated by state-dependent forces induced with laser beams \cite{Sorenson99,Sorenson2000,Milburn1999,Solano1999,Leibfried2003}. Combined with single-qubit gates and same-species entangling gates, this mixed-element entangling gate provides a complete set of gates over such a hybrid system for universal QIP \cite{Barenco1995,Bremner2002,Zhang2003}. Using a sequence of such gates, we demonstrate a Controlled-NOT (CNOT) gate and a SWAP gate \cite{NielsonChuang}. We further demonstrate the robustness of these gates against thermal excitation and show improved detection in quantum logic spectroscopy (QLS) \cite{Schmidt2005}. We also observe a strong violation of a CHSH-type Bell inequality \cite{CHSH1969} on entangled states composed of different ion species. }


Trapped ions of different elements vary in mass, internal atomic structure and spectral properties, features that can make certain species suited for particular tasks such as storing quantum information, high fidelity readout, fast logic gates, or interfacing between local processors and photon interconnects. One important advantage of a hybrid system incorporating trapped ions of different elements is the ability to manipulate and measure one type of qubit using laser beams with negligible effects on the other since the resonant transition wavelengths differ substantially. When scaling trapped-ion systems to greater numbers and density of ions, it will be advantageous to perform fluorescence detection on individual qubits without inducing decoherence on neighboring qubits due to uncontrolled photon scattering. To provide this function in a hybrid system one can use an entangling gate to transfer the qubit states to another ion species which is then detected without perturbing the qubits. This readout protocol could be further generalized to error correction schemes by extracting the error syndromes to the readout species while the computational qubits remain in the code. Another application could be in building photon interconnects between trapped-ion devices. Here, one species may be better suited for memory while the other is more favorable for coupling to photons \cite{Monroe2014,Moehring2007}. 

A mixed-element gate can also improve the readout in quantum logic spectroscopy (QLS) \cite{Schmidt2005}. In conventional quantum logic readout, the state of the clock or qubit ion is transferred to a motional state and in turn transferred to the detection ion, which is then detected with state-dependent fluorescence. In this case, the transfer fidelity directly depends on the purity of the motional state. In contrast, transfer utilizing the gate discussed here can be insensitive to the motion, as long as the ions are in the Lamb-Dicke regime \cite{bible}. This advantage extends to entanglement-assisted QND readout of qubit or clock ions, which can lower the overhead in time and number of readout ions as the number of clock ions increases \cite{Schulte2015}.

In our experiment, we use a beryllium ($^9$Be$^+$) ion and a magnesium ($^{25}$Mg$^+$) ion separated by approximately 4 $\mu$m along the axis of a linear Paul trap. The addressing lasers for each ion ($\lambda \simeq$ 313 nm for $^9$Be$^+$ and $\lambda \simeq$ 280 nm $^{25}$Mg$^+$) illuminate both ions. The qubits are encoded in hyperfine states of the ions. We choose $\left|F=2,m_F=0\right>=\left|\downarrow\right>_{\mathrm{Be}}$ and $\left|1,1\right>=\left|\uparrow\right>_{\mathrm{Be}}$ as the $^9$Be$^+$ qubit states and $\left|2,0\right>=\left|\downarrow\right>_{\mathrm{Mg}}$ and $\left|3,1\right>=\left|\uparrow\right>_{\mathrm{Mg}}$ for the $^{25}$Mg$^+$ qubit. The Coulomb coupling between the ions gives rise to two shared motional normal modes along the trap axis. A magnetic field of 11.945 mT is applied at 45 degrees with respect to the trap axis. At this field, the $^9$Be$^+$ qubit transition frequency is first-order insensitive to external magnetic field fluctuations \cite{Langer2005}. The magnetic field sensitivity of the $^{25}$Mg$^+$ qubit is approximately 430 kHz/mT. By measuring the decay of Ramsey interference fringes versus time between the Ramsey pulses on each qubit transition, we determine the $^9$Be$^+$ qubit's coherence time to be approximately 1.5 s. The $^{25}$Mg$^+$ qubit coherence time is approximately 6 ms limited by magnetic field fluctuations. We verified that the phase and contrast of Ramsey experiments on one species does not change measurably in the presence of light addressing the other species. This shows that the spectral separation is sufficient to isolate the species. 

Entanglement between the two ions is achieved through a M\o{}lmer-S\o{}rensen (MS) spin-spin interaction \cite{Sorenson99,Sorenson2000,Milburn1999,Solano1999} induced by laser-driven stimulated Raman transitions \cite{bible}. Starting in the state $\left|\uparrow\right>_{\mathrm{Be}}\left|\uparrow\right>_{\mathrm{Mg}}=\left|\uparrow\uparrow\right>$, the interaction can produce the Bell state $\Phi_{+}=\frac{1}{\sqrt{2}}\left(\left|\downarrow\downarrow\right>+\left|\uparrow\uparrow\right>\right)$ (see Methods). 

\begin{figure}
\includegraphics[width=1\linewidth]{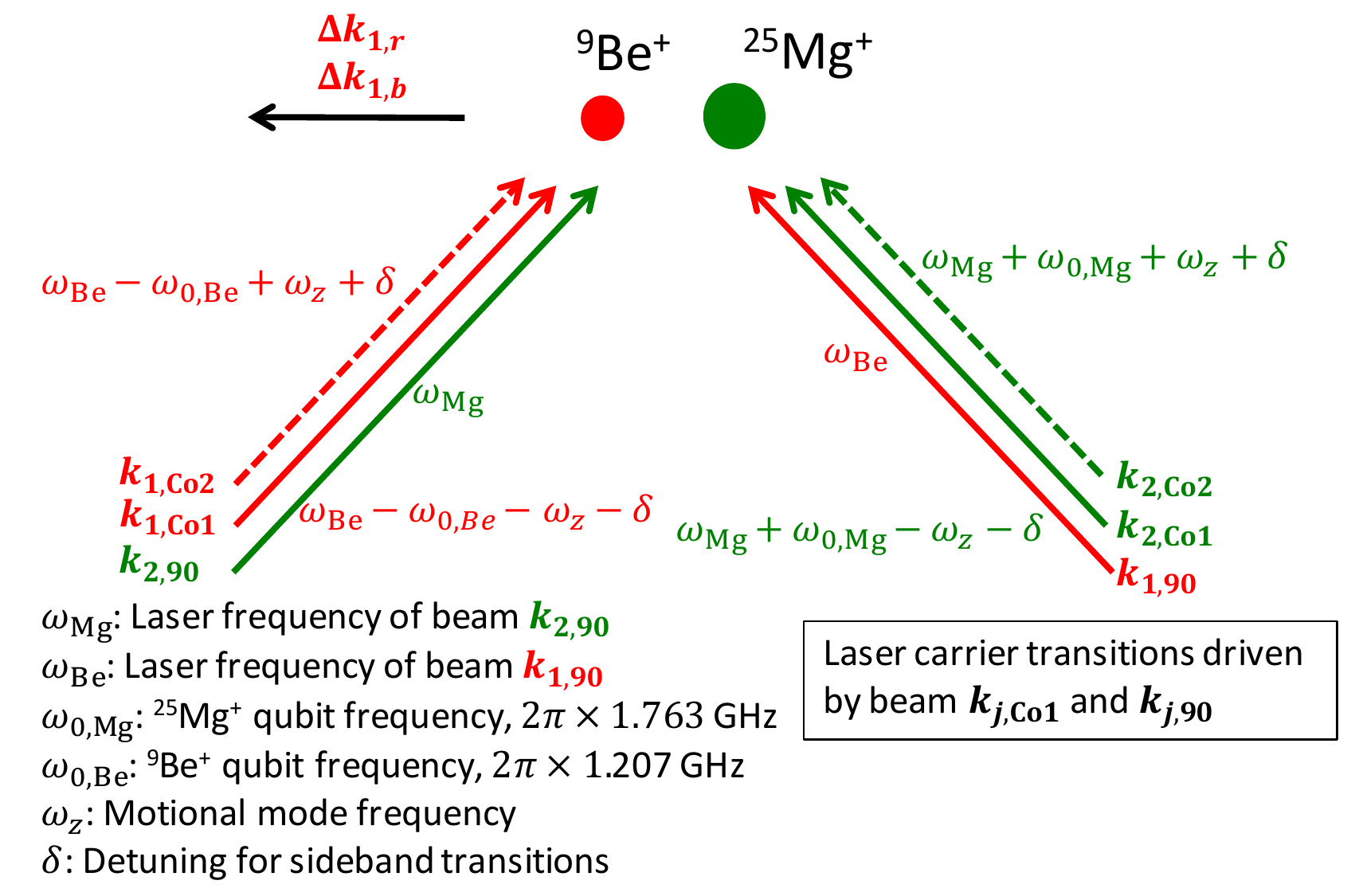}
\caption{
{\bf Configuration of laser beams for the mixed-element entangling gate.} For the $^9$Be$^+$ ion, 313 nm laser beams (in red) simultaneously induce near-resonant red and blue sidebands transitions. Similarly, for $^{25}$Mg$^+$, 280 nm beams (in green) induce sideband transitions. When all beams are applied simultaneously this implements the M\o{}lmer-S\o{}rensen spin-spin interaction (see Methods). Each set of qubit addressing laser beams is set up such that the wave vector differences $\bm{\Delta k_{j,r}}=\bm{k_{j,90}}-\bm{k_{j,\mathrm{Co1}}}$ and $\bm{\Delta k_{j,b}}=\bm{k_{j,90}}-\bm{k_{j,\mathrm{Co2}}}$ ($j=1,2$) are aligned in the same direction along the trap axis such that only motional modes along this axis can be excited. }
\label{BeamLine}
\end{figure}

The laser beam configurations to induce coherent Raman transitions are analogous for each element; for brevity, we will only describe the configuration for $^9$Be$^+$ (red in Fig. \ref{BeamLine}). Three laser beams, labeled by $\bm{k_{1,\mathrm{Co1}}}$, $\bm{k_{1,\mathrm{Co2}}}$ and $\bm{k_{1,90}}$, are derived from a single laser with wavelength $\lambda \simeq$ 313 nm. Beams $\bm{k_{1,\mathrm{Co1}}}$ and $\bm{k_{1,\mathrm{Co2}}}$ are copropagating such that their wave vector differences with respect to the $\bm{k_{1,90}}$ beam are aligned along the trap axis. In this configuration, only the axial motional modes interact with the laser beams. The two copropagating beams induce the detuned blue and red sideband Raman transitions, respectively, when paired with the $\bm{k_{1,90}}$ beam to implement the MS interaction (see Methods). 

\begin{figure*}
\centering
\includegraphics[width=1\linewidth]{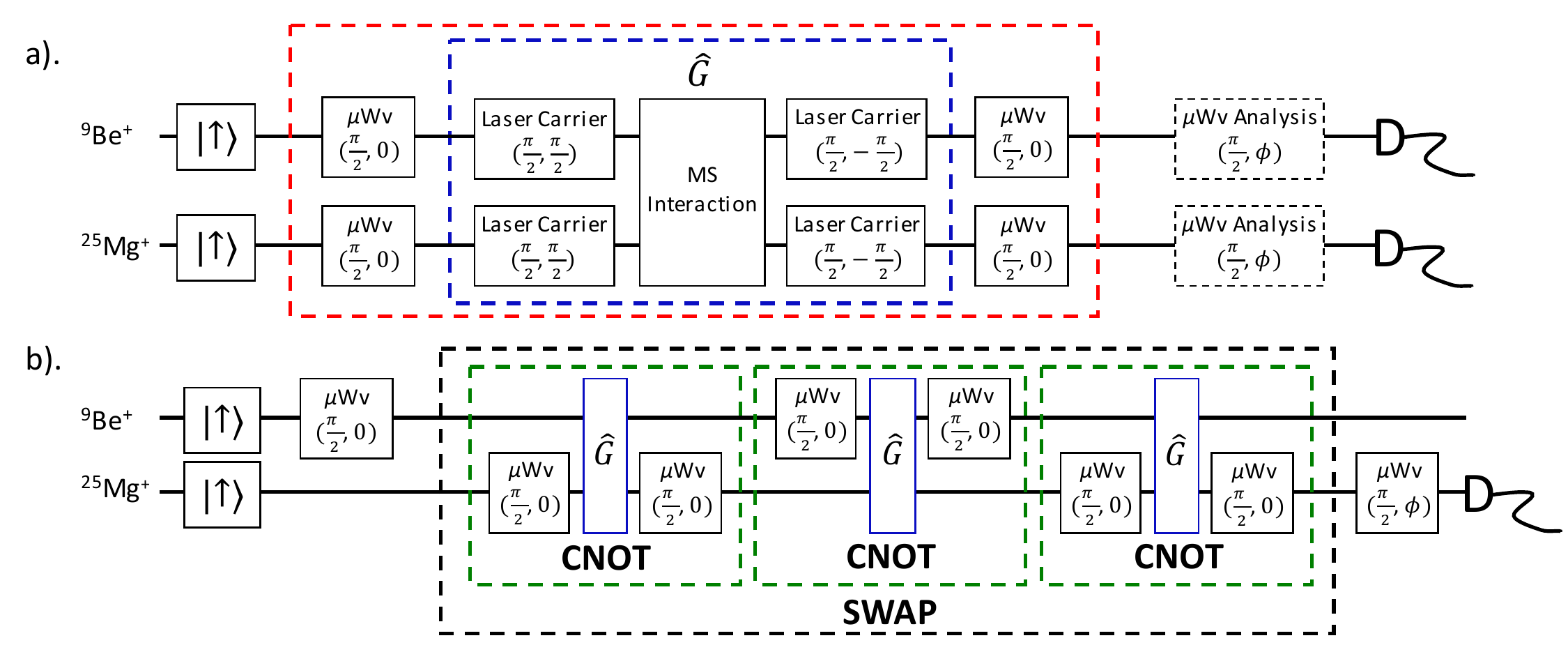}
\caption{
{\bf Pulse sequences for logics gates.} (a). Starting with the $\left|\uparrow\right>_{\mathrm{Be}}\left|\uparrow\right>_{\mathrm{Mg}}$ state, this pulse sequence generates a Bell state with $\widehat{G}$ (blue-dashed box) and single-qubit microwave ($\mu$Wv) gates. The notation $\left(\theta,\phi\right)$ represents the rotation angle and relative phase of each gate pulses. A parity oscillation is induced by applying analysis $\pi/2$ pulses with a variable phase $\phi$ to the created Bell state. To demonstrate the phase insensitivity of $\widehat{G}$, the single-qubit gates and the analysis pulses are implemented by microwave fields that are not phase synchronized to the optical phases. (b). Pulse sequence of a Ramsey experiment where a superposition state of a $^9$Be$^+$ qubit is coherently transferred to a $^{25}$Mg$^+$ qubit with a SWAP gate (black-dashed box). Given $\widehat{G}$, either of the two qubits can be the target qubit of a CNOT gate (green-dashed boxes) by applying single-qubit $\pi/2$ pulses to it. } 
\label{PulseSequence}
\end{figure*}

One important consideration in creating deterministic mixed-element entanglement with the MS interaction driven by multiple laser fields is the control over the relative optical phases at the ions' locations. The basis states $\left|+\right>_j$, $\left|-\right>_j$, and the state-dependent forces that are applied to them (see Methods) depend on the optical phases of the beams $\bm{k_{j,\mathrm{Co1}}}$, $\bm{k_{j,\mathrm{Co2}}}$, and $\bm{k_{j,90}}$ ($j=1,2$) at the ion positions. Beams $\bm{k_{j,\mathrm{Co1}}}$ and $\bm{k_{j,\mathrm{Co2}}}$ are generated in the same acousto-optic modulator (AOM), one for each ion species, and travel nearly identical paths. However, the $\bm{k_{j,90}}$ beams take a substantially different path to reach the ions' locations. Temperature drift and acoustic noise cause changes in the different beam paths that lead to phase fluctuations in the MS interaction. These fluctuations are slow on the timescale of a single gate but substantial over the course of many experiments. To suppress these effects, we embed the MS interaction in a Ramsey sequence implemented with two $\pi/2$ carrier pulses induced by $\bm{k_{j,\mathrm{Co1}}}$ and $\bm{k_{j,90}}$ for each qubit \cite{Lee05} (blue-dashed box in Fig. \ref{PulseSequence}.(a)). The first set of pulses maps the $\left|\uparrow\right>$ and $\left|\downarrow\right>$ states of each qubit onto the $\left|+\right>_j$ and $\left|-\right>_j$ states, whose phases are synchronized with the MS interaction. The final set of pulses undoes this mapping such that the action of this sequence is independent of the path length differences as long as the differences are constant during the entire sequence. In this case, the sequence produces a phase gate $\widehat{G}$ that implements $\left|\uparrow\uparrow\right>\rightarrow\left|\uparrow\uparrow\right>$, $\left|\uparrow\downarrow\right>\rightarrow i\left|\uparrow\downarrow\right>$, $\left|\downarrow\uparrow\right>\rightarrow i\left|\downarrow\uparrow\right>$, and $\left|\downarrow\downarrow\right>\rightarrow\left|\downarrow\downarrow\right>$. Such a phase gate could also be implemented as in Ref. [\citen{Leibfried2003}] (on qubits with magnetic-field-sensitive transitions). This requires fewer laser beams but adds the technical difficulty of synchronizing the state-dependent forces at the ion locations for both species. 

Before applying the gate, the ions are first Doppler cooled in all three directions. The axial motional modes are further cooled to near the ground state by Raman sideband cooling on the $^9$Be$^+$ ion \cite{Monroe1995}. State initialization into the qubits' $\left|\uparrow\right>$ states and qubit state readout are described in Methods. After each experiment repetition, we measure one of the possible states: $\left|\uparrow\uparrow\right>$, $\left|\uparrow\downarrow\right>$, $\left|\downarrow\uparrow\right>$, or $\left|\downarrow\downarrow\right>$. 

In a first experiment, we prepare the Bell state $\Phi_{+}$ with the MS interaction (Fig. 1) and determine its fidelity by measuring the qubit populations and the contrast of the parity oscillation by applying ``analysis'' pulses \cite{Sackett2000}. The analysis pulses are laser carrier transitions induced by the non-copropagating laser beams $\bm{k_{j,\mathrm{Co1}}}$ and $\bm{k_{j,90}}$ such that the relative phase defining the basis states of MS interaction is stable with respect to that of the analysis pulses for each experiment repetition. We determine a Bell state fidelity of $0.979(1)$. We also create a Bell state by applying microwave carrier $\pi/2$ pulses on each qubit before and after the operation $\widehat{G}$ (red-dashed box in Fig. \ref{PulseSequence}(a.)) achieving a fidelity of $0.964(1)$. Following the procedure of Ref. [\citen{Rowe2001}], we perform a CHSH-type Bell-inequality test \cite{CHSH1969} on this state achieving a sum of correlations of $B=2.70(2)>2$. This inequality, measured on an entangled system consisting of different elements, agrees with the predictions of quantum mechanics while eliminating the detection loophole but not the locality loophole \cite{Rowe2001}. 

The imperfections of the entangled states can be attributed to multiple causes which we investigate through calibration measurements and numerical simulation. We estimate the error from imperfect state preparation and detection to be $5\times10^{-3}$ (see Methods). Other errors are spontaneous photon scattering \cite{Ozeri2007} of $^{25}$Mg$^+$ ($6\times10^{-3}$) and $^9$Be$^+$ ($1\times10^{-3}$), and heating of the motional mode due to electric field noise ($4\times10^{-3}$) \cite{Turchette2000}. Other known error sources include imperfect single-qubit pulses, off-resonant coupling to spectator hyperfine states and the other motional modes, mode frequency fluctuations, qubit decoherence due to magnetic field fluctuations, laser intensity fluctuations, optical phase fluctuations, and calibration errors. Each of these sources contributes error on the order of $10^{-3}$ or less. We find close agreement between the experimental data and numerical simulations that include the listed imperfections. 

\begin{figure}
\includegraphics[width=1\linewidth]{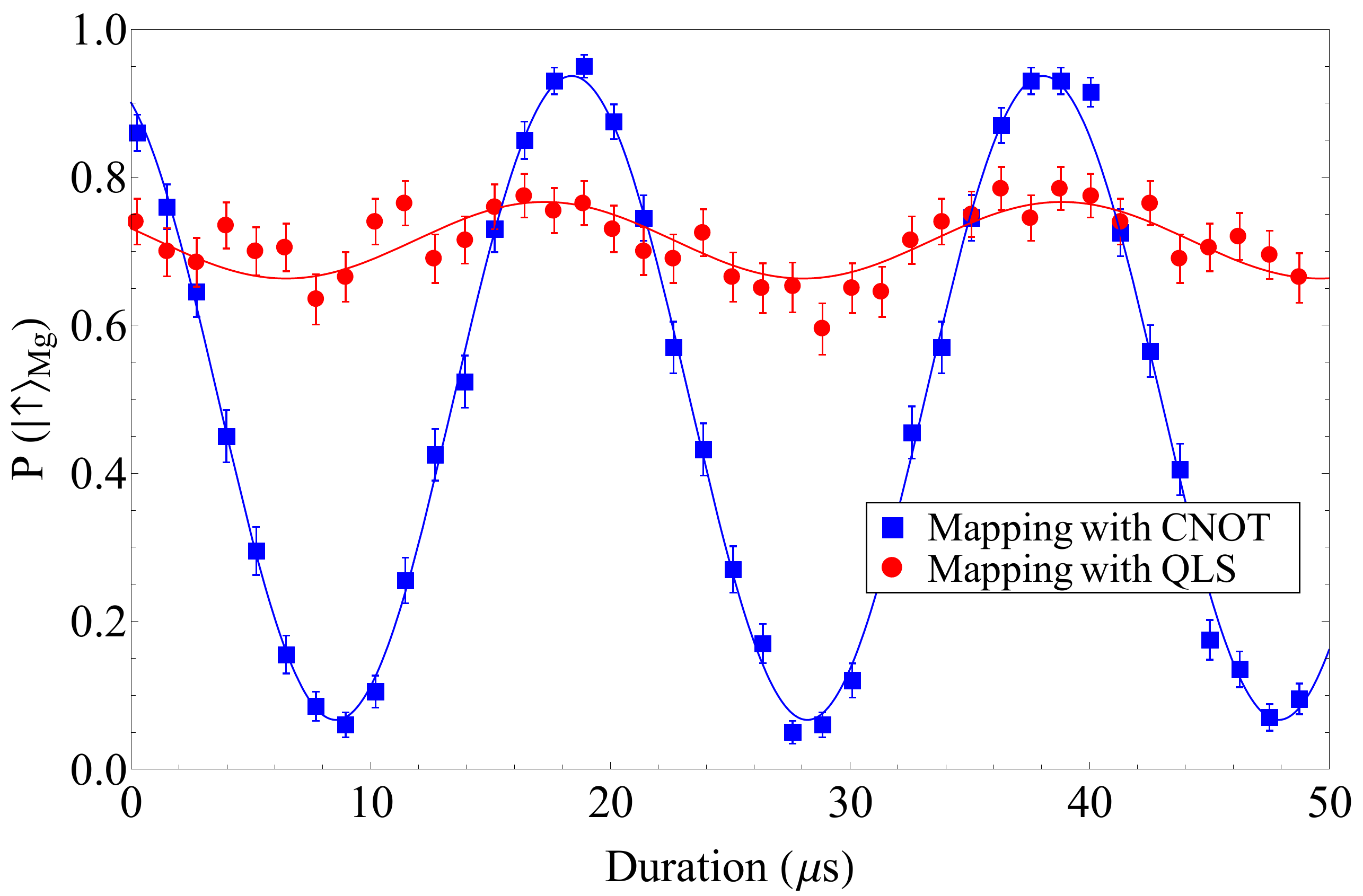}
\caption{
{\bf Robustness of quantum logic readout against thermal excitation.} Rabi flopping of the $^9$Be$^+$ ion detected on the $^{25}$Mg$^+$ ion with the motional modes cooled to Doppler temperatures using the two mapping procedures described in the text. P($\left|\uparrow\right>_{\mathrm{Mg}}$) is the probability of finding the $^{25}$Mg$^+$ qubit in the $\left|\uparrow\right>$ state. The CNOT mapping technique, which makes use of the mixed-species gate described here, performs better than the conventional QLS procedure due to the relative insensitivity to motional excitation. Each data represents $200$ repetitions and error bars correspond to s.e.m.}
\label{QLSContrast}
\end{figure}

\begin{figure}
\includegraphics[width=1\linewidth]{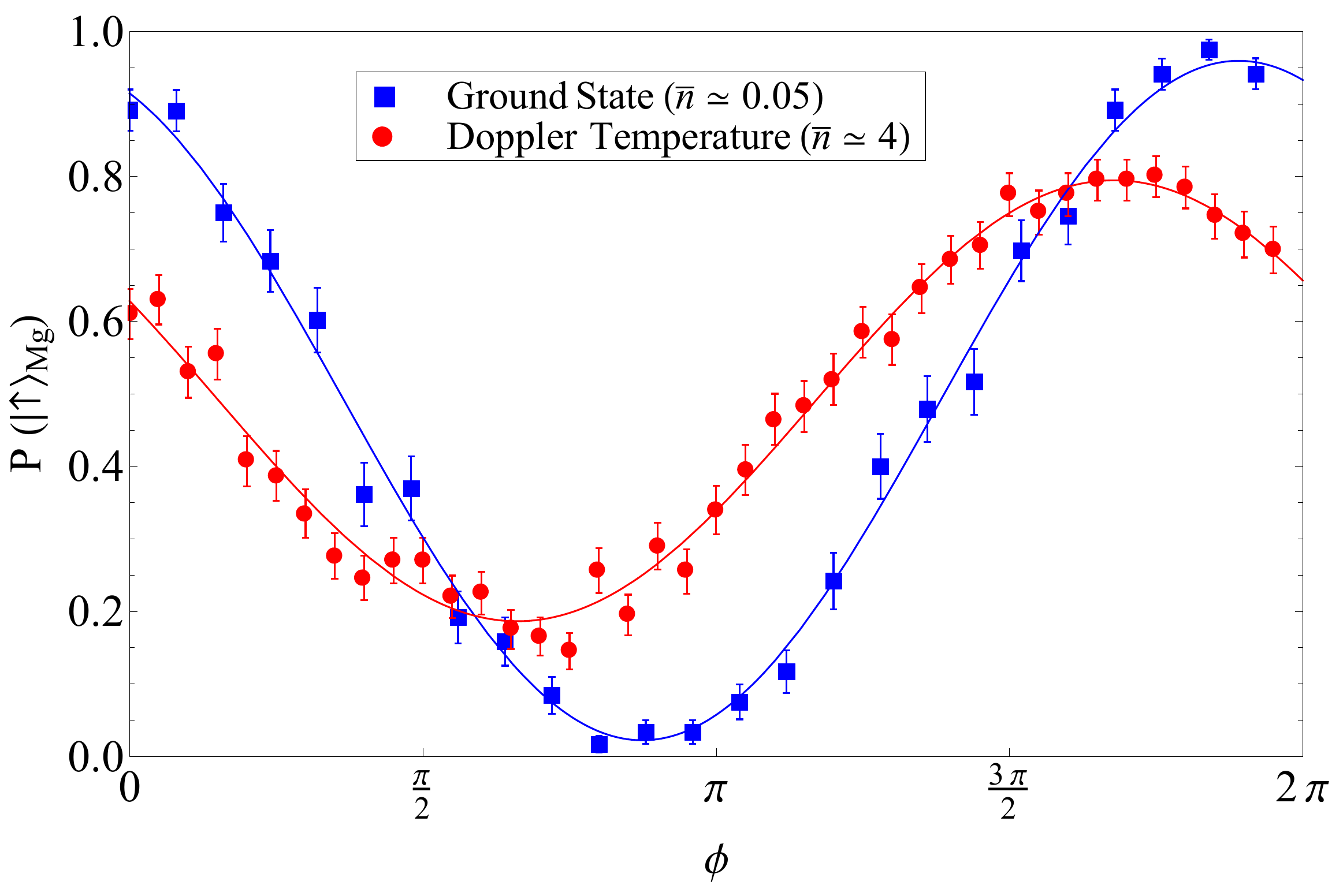}
\caption{
{\bf Ramsey experiments with SWAP gate.} Ramsey fringe of the $^{25}$Mg$^+$ qubit after initializing the $^9$Be$^+$ ion in the $\frac{1}{\sqrt{2}}\left(\left|\uparrow\right>+\left|\downarrow\right>\right)$ and applying the SWAP gate. P($\left|\uparrow\right>_{\mathrm{Mg}}$) is the probability of finding the $^{25}$Mg$^+$ qubit in the $\left|\uparrow\right>$ state and $\phi$ is the relative phase of the $\pi/2$ pulse applied to the $^{25}$Mg$^+$ qubit. The solid lines are fitted curves with contrast of 94 \% for the ions initialized to the ground state ($\bar{n}\simeq 0.05$) and 61 \% for the ions initialized to the Doppler cooling temperature ($\bar{n}\simeq 4$). The phase offset depends on the calibration of the SWAP gate and can be experimentally adjusted to any value. Error bars correspond to s.e.m. of 200 repetitions per data point. }
\label{SWAPGateContrast}
\end{figure}

We use $\widehat{G}$ to construct a CNOT gate by applying microwave $\pi/2$ pulses on one of the qubits before and after $\widehat{G}$ (green-dashed boxes in Fig. \ref{PulseSequence}(b.)) and use it to demonstrate qubit state mapping. The ``target'' of the CNOT gate is the qubit to which the single-qubit pulses are applied. The CNOT gate inherits the robustness against motional excitation from the MS gate \cite{Sorenson99,Sorenson2000,Milburn1999,Solano1999}. We compare the CNOT gate with the method used in conventional QLS procedure where a red-sideband pulse is first applied to the $^9$Be$^+$ ion followed by a red-sideband pulse to the $^{25}$Mg$^+$ ion \cite{Schmidt2005}. Both procedures are calibrated for the motional mode ground state. Figure \ref{QLSContrast} shows Rabi flopping of the $^9$Be$^+$ qubit as detected on the $^{25}$Mg$^+$ ion, which is initially prepared in the $\left|\uparrow\right>$ state. For the ions' motional modes cooled to Doppler temperature (mean occupation number $\bar{n}\simeq 4$), the contrast of conventional QLS method (red dots) is reduced significantly compared to transfer with CNOT gate (blue squares). In both of these mapping procedures the $^9$Be$^+$ qubit phase information is not accessible on the $^{25}$Mg$^+$ ion. To preserve this phase information, we construct a SWAP gate that interchanges the quantum state of the two qubits \cite{NielsonChuang} with three CNOT gates. Figure \ref{PulseSequence}.(b) shows the pulse sequence of a Ramsey-type experiment where the first Ramsey (microwave) $\pi/2$ pulse is applied to the $^9$Be$^+$ ion and the second (microwave) $\pi/2$ pulse is applied to the $^{25}$Mg$^+$ ion after implementing the SWAP gate. Ramsey fringes for the ions' axial motional modes initialized to near the ground state ($\bar{n}\simeq 0.05$, blue squares) and Doppler cooled ($\bar{n}\simeq 4$, red dots) are shown in Fig. \ref{SWAPGateContrast}. The contrast at Doppler temperature is reduced because the Lamb-Dicke limit is not rigorously satisfied. Through simulation with and without the measured $^{25}$Mg$^+$ qubit decoherence, we determine that the loss of contrast for the SWAP gate due to this decoherence is approximately 2 \%. For all three methods, the contrasts could be somewhat improved by calibrating all gates for the given motional temperature. 

In summary, we have demonstrated a mixed-element entangling gate where we employ a Ramsey sequence to suppress loss of fidelity of the output state due to low-frequency optical path length fluctuations \cite{Lee05}. Using this gate, we implement CNOT and SWAP operations between qubit elements which are relatively robust against thermal excitation of the motion. These and related techniques are potentially useful for building a large scale processor or quantum network utilizing the advantageous properties of different ion species \cite{MonroeandKim2013,Monroe2014}. The entangling technique should also be applicable to qubits with optical transitions (e.g. Ca$^+$ ion or Sr$^+$ ion), or a combination of hyperfine qubits and optical qubits, which can also make this technique useful for readout in quantum logic clocks \cite{Chou2010}. 

Similar work has also been carried out at the University of Oxford \cite{Ballance2015} on different isotopes of Ca$^+$ where the same laser beams can manipulate both isotopes simultaneously. The method presented here uses two substantially different sets of laser beams and ion spectra illustrating that cross-talk between operations on different species can be negligible, and could be applied to take advantage of each species' desirable features. 

\section{Methods}

\subsection{Geometric phase gate}

The M\o{}lmer-S\o{}rensen (MS) protocol \cite{Sorenson99,Sorenson2000,Milburn1999,Solano1999} requires simultaneous excitation of a blue sideband transition with a detuning of $\delta$ and a red sideband transition with a detuning of $-\delta$ for a selected motional mode (Fig. \ref{BeamLine}). The excitation creates a forced harmonic oscillator interaction and displaces the motional wavefunction in phase space in a manner that is dependent on the internal qubit states. If the different displacements enclose a loop, the qubit states pick up a geometric phase proportional to the state-dependent area of the enclosed loop. We create an entangling logic gate by choosing appropriate geometric phase differences between different qubit states. 

Laser fields are used to induce coherent stimulated-Raman transitions between the qubit states of each ion and the shared quantized degrees of motion \cite{bible}. For each qubit we can excite carrier transitions $|\downarrow ,n \rangle \leftrightarrow |\uparrow ,n \rangle$ that induce spin-flips without changing the motional Fock state $n$. A blue (red) sideband excitation $|\downarrow ,n \rangle \leftrightarrow |\uparrow ,n + 1\rangle$  ($|\uparrow ,n - 1\rangle$) flips the spin while adding (removing) a quantum of motion by detuning the fields from the carrier transition frequency by the motional frequency in the positive (negative) direction. The relative frequencies, phases, and intensities of each set of laser beams (Fig. \ref{BeamLine}) for each qubits can be adjusted with AOMs which are computer controlled. The $^9$Be$^+$ Raman laser beams with a wavelength of $\lambda \simeq$ 313 nm are approximately $~480$ GHz red detuned from the S$_{1/2}$ to P$_{1/2}$ electronic state transition. The $\lambda \simeq$ 280 nm Raman laser beams for $^{25}$Mg$^+$ ion are approximately $~160$ GHz blue detuned from the S$_{1/2}$ to P$_{3/2}$ electronic state transition. Carrier transitions can also be implemented by microwave fields delivered from an antenna located outside the vacuum chamber. 

After transforming into the respective interaction frames of both qubits as well as that of the shared motional mode of motion, and dropping high-frequency terms in the rotating-wave approximation, we can write the interaction in the Lamb-Dicke limit as \cite{bible}
\begin{eqnarray}
\nonumber H = \hbar\displaystyle\sum_{j=1,2} \mathit{\Omega}_j \hat{\sigma}_j^\dagger \left(\hat{a} \mathrm{e}^{-\mathrm{i}(\delta_j t - \phi_{j,r})} +\hat{a}^\dagger \mathrm{e}^{\mathrm{i}(\delta_j t + \phi_{j,b})}\right)+h.c.,
\label{eqnZbasis}
\end{eqnarray}
where $j=1,2$ denotes the two different ion species, $\mathit{\Omega}_j = \eta_j \mathit{\Omega}_{0,j}$ where $\mathit{\Omega}_{0,j}$ is the carrier resonance Rabi frequency. The Lamb-Dicke parameter $\eta_j$ is equal to $\Delta k_j z_{0,j} b_j$, where $b_j$ is the mode amplitude of the {\it j}th ion and $z_{0,j}=\sqrt{\hbar/2 m_j \omega_{z}}$, $m_j$ is the mass and $\omega_{z}$ is the frequency of the selected normal mode. The spin raising operator is $\hat{\sigma}_j^\dagger$ and $\hat{a}^\dagger$ is the creation operator for the relevant (harmonic) motional mode. 

The phases of the red (r) and blue (b) sideband interactions are $\phi_{j,r(b)} = \Delta k_{j,r(b)} X_{0,j} + \Delta\phi_{j,r(b)}$ where $\Delta k_{j,r(b)}$ and $\Delta\phi_{j,r(b)}$ are the differences in wave vectors and phases of the optical fields driving the red and blue sideband transitions respectively, and $X_{0,j}$ is the equilibrium position for the {\it j}th ion. After setting $\mathit{\Omega}_1=\mathit{\Omega}_2=\mathit{\Omega}$ and $\delta_1=\delta_2=\delta$, and writing $\phi_{M,j}=\left(\phi_{j,r}-\phi_{j,b}\right)/2$, the geometric phases accumulated after a duration of $t_{\mathrm{MS}}=2\pi/\delta$ for the four $\left|+\right>_j$ and $\left|-\right>_j$ basis states (defined as the eigenstates of $\hat{\sigma}_{\phi,j} = \cos\left((\phi_{j,r}+\phi_{j,b})/2\right)\hat{\sigma}_{x,j}-\sin\left((\phi_{j,r}+\phi_{j,b})/2\right)\hat{\sigma}_{y,j}$) are
\begin{eqnarray}
\nonumber \varphi_{\left|+,+\right>,\left|-,-\right>} &=& \frac{8\pi\mathit{\Omega}^2}{\delta^2}\cos^2\left(\frac{\phi_{M,1}-\phi_{M,2}}{2}\right),\\
\varphi_{\left|+,-\right>,\left|-,+\right>} &=& \frac{8\pi\mathit{\Omega}^2}{\delta^2}\sin^2\left(\frac{\phi_{M,1}-\phi_{M,2}}{2}\right).
\label{GeometricPhase}
\end{eqnarray}
To maximize entangling gate speed, the geometric phases for the different parity qubit states in Eq. \ref{GeometricPhase} are set to differ by $\pi/2$. This is accomplished by adjusting the phases of the radio frequencies driving the AOMs. 

There are two axial modes: the lower frequency mode ($\omega_z = 2\pi\times2.5$ MHz), where the ions oscillate in phase, and the higher frequency mode ($2\pi\times5.4$ MHz), where the ions oscillate out of phase. The Lamb-Dicke parameter for the $^9$Be$^+$ ($^{25}$Mg$^+$) ion is 0.156 (0.265) and 0.269 (0.072), respectively, for the two modes. We use the in-phase mode for our demonstration because for the $^{25}$Mg$^+$ ion, it has a larger normal mode amplitude compared to the out-of-phase mode. This results in less spontaneous emission error for a given strength of the state-dependent force. Gate time $t_{\mathrm{MS}}$ is approximately 35 $\mu$s.

\subsection{Calibration procedure for phase gate $\widehat{G}$}

To produce the phase gate $\widehat{G}$, the phases of the $\pi/2$ pulses for the Ramsey sequence must be referenced to the basis states of the MS interaction defined by the optical phases. The phases must also account for the AC Stark shifts induced by the laser beams that are used for the MS interaction. 

To calibrate these phases, we first perform the pulse sequence shown in the blue-dashed box of Fig. \ref{PulseSequence}.(a) with the MS interaction pulses detuned far off-resonant from the red and blue sideband transitions such that they only induce AC Stark shifts on the qubits. Starting with the input state $\left|\uparrow\uparrow\right>$, we set the phases of the final $\pi/2$ laser pulses such that the action of this pulse sequence returns each qubit to the $\left|\uparrow\right>$ state. Then, we perform this sequence with the MS interactions correctly tuned and vary the phases of the MS interactions. Again, in this case we look for the phase that maps the input state $\left|\uparrow\uparrow\right>$ back to itself. We verify the action of this $\widehat{G}$ operation by creating a Bell state with the pulse sequence shown in Fig. \ref{PulseSequence}.(a). 

\subsection{Qubit state preparation and readout}

For qubit state preparation, the $^9$Be$^+$ ion is optically pumped to the $\left|2,2\right>$ state followed by Doppler cooling implemented by driving the S$_{1/2}\left|2,2\right>\leftrightarrow$ P$_{3/2}\left|3,3\right>$ cycling transition with a $\sigma^+$ polarized light. Similarly, we optically pump the $^{25}$Mg$^+$ ion to the $\left|3,3\right>$ state and apply Doppler cooling on the S$_{1/2}\left|3,3\right>\leftrightarrow$ P$_{3/2}\left|4,4\right>$ transition. For ground state initialization of the axial motional modes, Raman sideband cooling is applied to the $^9$Be$^+$ ion \cite{Monroe1995}. To transfer the $^9$Be$^+$ $\left|2,2\right>$ state to the $\left|1,1\right>=\left|\uparrow\right>_{\mathrm{Be}}$ state, we use microwave composite pulse sequences that are robust against transition detuning errors. These consist of resonant $\left(\frac{\pi}{2},0\right)$, $\left(\frac{3\pi}{2},\frac{\pi}{2}\right)$, $\left(\frac{\pi}{2},0\right)$ pulses \cite{Levitt1986}, where the first entry denotes the angle the state is rotated about a vector in the {\it x-y} plane of the Bloch sphere and the second angle represents the azimuthal angle for the rotation axis. With analogous sequences, we first transfer the $^{25}$Mg$^+$ from the $\left|3,3\right>$ state to the $\left|2,2\right>$ state, and then to the $\left|3,1\right>=\left|\uparrow\right>_{\mathrm{Mg}}$ state. 

The state-dependent resonance-fluorescence detection technique is accomplished with an achromatic lens system designed for 313 nm and 280 nm \cite{Huang2004}. We sequentially image each ion's fluorescence onto a photomultiplier tube. After reversing the initial mapping procedures to put the $\left|\uparrow\right>$ states back in the respective cycling transition ground states, we apply the Doppler cooling beams. The fluorescing or ``bright'' state of this protocol therefore corresponds to the $\left|\uparrow\right>$ state of each ion. The $\left|\downarrow\right>$ state of each qubit is transferred to $\left|1,-1\right>$ and $\left|2,-2\right>$ for the $^9$Be$^+$ and $^{25}$Mg$^+$, respectively, with microwave carrier $\pi$ pulses. These states are ``dark'' to the detection beams and correspond to the $\left|\downarrow\right>$ state. This ``shelving'' technique is used to minimize the overlap of the bright and dark state photon count probability distributions. With detection durations of 330 $\mu$s for $^9$Be$^+$ and 200 $\mu$s for $^{25}$Mg$^+$, we detect on average 30 photons for each ion when they are in the bright state and 3.5 photons (predominantly from background light) when they are in the dark state. The qubit state is determined by choosing a photon count threshold such that the states are maximally distinguished. The state preparation and detection error of $5\times10^{-3}$ reported in the main text includes errors due to the threshold detection protocol (false determination of each detected state being in the other state) and the infidelities of the microwave transfer pulses. 

\begin{acknowledgments}
We thank J. Bollinger and D. Hume for helpful comments on the manuscript. This work was supported by the Office of the Director of National Intelligence (ODNI) Intelligence Advanced Research Projects Activity (IARPA), ONR, and the NIST Quantum Information Program. Y. Wan is supported by the U.S. Army Research Office through MURI grant W911NF-11-1-0400. This paper is a contribution by NIST and not subject to U.S. copyright. 
\end{acknowledgments}

\section{Author Contributions}
T.R.T. and J.P.G. conceived and designed the experiments, developed components of the experimental apparatus, collected and analyzed data. T.R.T. wrote the manuscript. Y.L., Y.W., and R.B. contributed to the development of experimental apparatus. D.L. and D.J.W. directed the experiment. All authors provided important suggestions for the experiments, discussed the results, and contributed to the editing of the manuscript. 

\section{Author Information}
The authors declare no competing financial interests. Correspondence and requests for materials should be addressed to T.R.T. (tingrei.tan@nist.gov).


\begin{thebibliography}{1}

\bibitem{Wallquist2009}
Wallquist, M., Hammerer, K., Rabl, P., Lukin, M., Zoller, P. Hybrid quantum devices and quantum engineering. {\it Phys. Scr.} {\bf T137}, 014001 (2009).

\bibitem{Barrett2003}
Barrett, M. D. {\it et al.} Sympathetic cooling of $^9$Be$^+$ and $^{24}$Mg$^+$ for quantum logic. {\it Phys. Rev. A} {\bf 68}, 042302 (2003).

\bibitem{Lin2013}
Lin, Y. {\it et al.} Dissipative production of a maximally entangled steady state of two quantum bits. {\it Nature} {\bf 504}, 415-418 (2013).

\bibitem{Hume2007}
Hume, D. B. {\it et al.} High-fidelity adaptive qubit detection through repetitive quantum nondemolition measurements. {\it Phys. Rev. Lett.} {\bf 99},  120502 (2007).

\bibitem{Sorenson99}
S\o{}rensen, A. \& M\o{}lmer, K. Quantum computation with ions in thermal motion. {\it Phys. Rev. Lett.} {\bf 82}, 1971-1974 (1999).

\bibitem{Sorenson2000}
S\o{}rensen, A. \& M\o{}lmer, K. Entanglement and quantum computation with ions in thermal motion. {\it Phys. Rev. A} {\bf 62}, 022311 (2000).

\bibitem{Milburn1999}
Milburn, G. J., Schneider, S., James, D. F. V. Ion trap quantum computing with warm ions. {\it Fortschr. Phys.} {\bf 48}, 801-810 (2000).

\bibitem{Solano1999}
Solano, E., de Matos Filho, R. L., Zagury, N. Deterministic Bell states and measurement of the motional state of two trapped ions. {\it Phys. Rev. A} {\bf 59},  R2539-R2543 (1999).

\bibitem{Leibfried2003}
Leibfried, D. {\it et al.} Experimental demonstration of a robust, high-fidelity geometric two ion-qubit phase gate. {\it Nature} {\bf 422}, 412-415 (2003).

\bibitem{Barenco1995}
Barenco, A. {\it et al.} Elementary gates for quantum computation. {\it Phys. Rev. A} {\bf 52}, 3457-3467 (1995).

\bibitem{Bremner2002}
Bremner, M. J. {\it et al.} Practical scheme for quantum computation with any two-qubit entangling gate. {\it Phys. Rev. Lett.} {\bf 89}, 247902 (2002).

\bibitem{Zhang2003}
Zhang, J. Vala, J. Sastry, S. \& Whaley, K. B. Exact two-qubit universal quantum circuit. {\it Phys. Rev. Lett.} {\bf 91}, 027903 (2003).

\bibitem{NielsonChuang}
Nielson, M. A. \&  Chuang, I. L. Quantum Computation and Quantum Information (Cambridge Univ. Press, Cambridge, 2000).

\bibitem{Schmidt2005}
Schmidt, P. O. {\it et al.} Spectroscopy using quantum logic. {\it Science} {\bf 309}, 749-752  (2005).

\bibitem{CHSH1969}
Clauser, J. F., Horne, M. A., Shimony, A., Holt, R. A. Proposed experiment to test local hidden-variable theories. {\it Phys. Rev. Lett.} {\bf 23}, 880-884 (1969).

\bibitem{Monroe2014}
Monroe, C. {\it et al.} Large-scale modular quantum-computer architecture with atomic memory and photonic interconnects. {\it Phys. Rev. A} {\bf 89}, 022317 (2014).

\bibitem{Moehring2007}
Moehring, D. L. {\it et al.} Quantum networking with photons and trapped atoms. {\it JOSA B} {\bf 24}, 300-315 (2007)


\bibitem{bible}
Wineland, D. J. {\it et~al.}, {\it J. Res. Natl. Inst. Stand. Technol.} {\bf 103},  259-328 (1998).

\bibitem{Schulte2015}
Schulte, M., L\"{o}rch, N. , Leroux, I. D., Schmidt, P. O., Hammerer, K. Quantum algorithmic readout in multi-ion clocks. arXiv:1501.06453 (2015).

\bibitem{Langer2005}
Langer, C. {\it et~al.} Long-lived qubit memory using atomic ions. {\it Phys. Rev. Lett.} {\bf 95}, 060502 (2005).

\bibitem{Lee05}
Lee, P. J. {\it et al.} Phase control of trapped ion quantum gates. {\it J. Opt. B} {\bf 7}, S371-S383 (2005).

\bibitem{Monroe1995}
Monroe, C. {\it et al.} Resolved-sideband Raman cooling of a bound atom to the 3D zero-point energy. {\it Phys. Rev. Lett.} {\bf 75}, 4011-4014 (1995).

\bibitem{Sackett2000}
Sackett, C. A. {\it et al.} Experimental entanglement of four particles. {\it Nature} {\bf 404}, 256-259 (2000).

\bibitem{Rowe2001}
Rowe, M. A. {\it et al.} Experimental violation of a Bell's inequality with efficient detection. {\it Nature} {\bf 409}, 791-794 (2001).

\bibitem{Ozeri2007}
Ozeri, R. {\it et~al.} Errors in trapped-ion quantum gates due to spontaneous photon scattering. {\it Phys. Rev. A} {\bf 75}, 042329 (2007).

\bibitem{Turchette2000}
Turchette, Q. A. {\it et~al.} Heating of trapped ions from the quantum ground state. {\it Phys. Rev. A} {\bf 61}, 063418 (2000).

\bibitem{MonroeandKim2013}
Monroe, C. \& Kim, J. Scaling the ion trap quantum processor. {\it Science} {\bf 339},  1164-1169  (2013).

\bibitem{Chou2010}
Chou, C. W., Hume, D. B., Koelemeij, J. C. J, Wineland, D. J., Rosenband, T. Frequency comparison of two high-accuracy Al$^+$ optical clocks. {\it Phys. Rev. Lett.} {\bf 104}, 070802 (2010).

\bibitem{Ballance2015}
Ballance, C. J. {\it et al.} Hybrid quantum logic and a test of Bell's inequality using two different atomic species. arXiv:1505:04014 (2015).

\bibitem{Levitt1986}
Levitt, M. H. Composite Pulses. {\it Prog. NMR Spectrosc.} {\bf 18}, 61-122 (1986).

\bibitem{Huang2004}
Huang, P. \& Leibfried D. Achromatic catadioptric microscope objective in deep ultraviolet with long working distance. {\it Proc. SPIE} {\bf 5524}, 125-133 (2004).

\end{thebibliography}
\end{document}